\documentclass[12pt,preprint]{aastex}

\newcommand{\kms}  {km s$^{-1}$}
\newcommand{\msol} {M$_{\odot}$}
\def\lesssim{\mathrel{\hbox{\rlap{\hbox{\lower4pt\hbox{$\sim$}}}\hbox{$<$}}}}
\def\gtrsim{\mathrel{\hbox{\rlap{\hbox{\lower4pt\hbox{$\sim$}}}\hbox{$>$}}}}

\slugcomment{Submitted for publication in ApJ}

\shorttitle{The progenitor of SN1999em}
\shortauthors{Smartt et al.}

\begin{document}


\title{The nature of the progenitor of the Type\,II-P supernova 
1999em\thanks{Based on observations at the William Herschel Telescope
on La Palma and the Canada-France-Hawaii Telescope}}


\author{Stephen J. Smartt,  Gerard F. Gilmore, Christopher A. Tout, 
Simon T. Hodgkin}
\affil{Institute of Astronomy, University of Cambridge, Madingley Road,
       CB3 OHA, Cambridge, England}


\email{sjs@ast.cam.ac.uk}


\begin{abstract}
The masses and the evolutionary states of the progenitors of
core-collapse supernovae are not well constrained by direct
observations.  Stellar evolution theory generally predicts that
massive stars with initial masses less than about 30\msol\ should
undergo core-collapse when they are cool M-type supergiants. However
the only two detections of a SN progenitor before explosion are
SN1987A and SN1993J, and neither of these was an M-type
supergiant. Recently we have set an upper mass-limit to the progenitor
of Type II-P SN1999gi of 9$^{+3}_{-2}$\msol\ (Smartt et al. 2001a) from
pre-explosion Hubble Space Telescope images. In this paper we present
high quality ground-based $VRI$ images of the site of the Type II-P
SN1999em (in NGC1637) taken before explosion,  which were extracted
from the CFHT archive.  We determine a precise position of the SN on
these images to an accuracy of $0.17''$. The host galaxy is close
enough ($7.5\pm0.5$\,Mpc) that the bright supergiants are resolved as
individual objects, however we show that there is no detection of an
object at the SN position before explosion that could be interpreted
as the progenitor star. By determining the sensitivity limits of the
$VRI$ data, and assuming the reddening toward the progenitor is
similar to that toward the SN itself, we derive bolometric luminosity
limits for the progenitor. 
Comparing these to standard stellar evolutionary tracks which trace
evolution up to the point of core carbon ignition, we initially 
derive an upper mass limit of 
approximately 12\msol. However we present evolutionary 
calculations that follow 7-12\msol\ stars throughout their C-burning lifetime
and show that we can restrict the mass of the progenitor even further. 
Our calculations indicate that progenitors initially of 8-10\msol,  
undergoing expected mass loss, can also be excluded
because a second dredge up sends them to
somewhat higher luminosities than a star of initially $12\,M_\odot$.
These results limit the progenitor's initial main-sequence
mass to a very narrow range of 12$^{+1}_{-1}$\msol. 
We discuss the similarities between the Type II-P SNe 1999em and 1999gi
and their progenitor mass limits, and also consider all the direct
evidence currently available on progenitors of core-collapse SNe. 
We suggest that SN Type II-P originate only in intermediate mass stars of
8-12\msol, which are in the red supergiant region and that higher mass
stars produce the other Type\,II sub-types. 
Finally we present a 
discussion on the future possibilities of determining masses, or
mass-limits, and evolutionary status of core-collapse events on a
statistically larger sample, using virtual observatory initiatives
such as {\sc astrovirtel}. 
\end{abstract}



\section{Introduction}
%
Supernovae are the evolutionary end points of all stars more massive
than about 8\msol. Predictions of the pre-explosion evolutionary
status of these stars is a key test of stellar evolutionary theory.
Supernovae explosions additionally drive the chemical evolution of the
Universe and play a major role in shaping the dynamics of the
interstellar medium of gas rich galaxies. They are of crucial
importance in the fundamental studies of the evolution of galaxies and
the origins of the chemical elements in the Universe.

The spectra of supernovae come
in many different varieties with the classifications based on the lines
observed and the temporal evolution of these features. The presence of
broad H\,{\sc i} optical lines indicates a SN Type\,II
classification, while those that do not show hydrogen are classed
Type\,I. The SNe\,Ia are thought to arise through thermonuclear
explosions in white dwarf binary systems, hence the progenitors are
low-intermediate mass stars. All other supernovae including the Type
Ib/Ic and all flavours of Type II are thought to be due to core-collapse
during the deaths of massive stars. SNe\,II show
prominent, broad H\,{\sc i} lines in their optical spectra, indicating
that the progenitor retained a substantial hydrogen envelope prior to
explosion. SNe\,Ib/Ic do not show any significant signs of hydrogen
in their spectra, although SNe Ib display pronounced He\,{\sc i}
absorption. There is very strong evidence that the SNe II and Ib/Ic
are associated with the deaths of massive stars as they are never seen
in elliptical galaxies, only rarely in S0 types and they often appear
to be associated with sites of recent starformation such as H\,{\sc
ii} regions and OB associations in spiral and irregular galaxies 
\citep{vandyk96,fili97}.  The Type\,II
events are further seen to be split into subtypes (IIb, IIn, II-L and
II-P). \citet{leo2001} discuss the widely held belief
that core-collapse events can be ranked in order of their increasing
hydrogen envelope mass at the time of explosion, which is - Ic, Ib,
IIb, IIn, II-L, II-P.

This overwhelming, but still indirect, evidence implies that SNe\,II
arise from the deaths of single, massive stars with initial masses M$>
8-10$M$_{\odot}$ and which have retained a substantial fraction of
their hydrogen envelope.  However there has been only one definite and
unambiguous detection of a star that has subsequently exploded as a
SNe of any type $-$ that of Sk$-69^{\circ}202$, the progenitor to
SN1987A in the LMC \citep{white87}. Prior to explosion
this star was a blue supergiant of B3\,Ia spectral type \citep{wal89}, 
which would correspond to $T_{\rm eff} =18000$\,K
\citep[from the temperatures in][]{mcer99} and $\log
L/L_{\odot} = 5.1$ \citep[from the photometry in][]{wal89}, 
and an initial mass of $\sim$20M$_{\odot}$. The
closest supernova to the Milky Way since then was SN1993J in M81 (3.63\,Mpc), 
which was a Type\,IIb event and  
ground based $UBVRI$ photometry of the SN site before explosion
was presented by \citet{alder94}. The photometry of
the progenitor candidate was best fit with a composite spectral energy
distribution of a K0\,Ia star and some excess $UB$ band flux either
from unresolved OB association contamination or a hot
companion. Neither the progenitor of SN1987A nor that of SN1993J is consistent
with the canonical stellar evolution picture, where core carbon
burning finishes and core-collapse occurs relatively soon afterwards
($\sim10^3-10^4$\,yrs) while the massive star is an
M-supergiant. 

Other attempts have been made to directly find SNe
progenitors on pre-explosion archive images, with little success in
directly detecting progenitor stars.  An upper mass limit to the
progenitor of SN1980K has been estimated to be $\sim$18M$_{\odot}$
\citep{thom82}, while only an upper limit to the absolute visual
magnitude was determined for SN1994I \citep{barth96}. 
Recently \citet{smartt2001}
studied HST archive images of the site of the Type\,II-P SN1999gi
which were taken before explosion.  The SN1999gi occurred in a young
OB-association, however the the progenitor was below the detection
limit of the two pre-explosion images. By determining the sensitivity
of these exposures and comparing the estimated bolometric luminosity
with stellar evolutionary theory, an upper limit to the mass of the
progenitor was set at 9$^{+3}_{-2}$\msol.

The SN1999em was discovered on Oct. 29 1999 by the Lick 
Observatory Supernova Search in NGC1637 \citep{li99} at an
unfiltered CCD magnitude of $\sim13.5^m$. It was 
soon confirmed to be a Type\,II and being a very 
bright event it has been studied extensively in the 
optical since then. It has been firmly established as 
a normal Type II-P event, having a plateau phase lasting 
approximately 90 days after discovery \citep{leo2001}. 
There have also been UV, X-ray, radio, and spectropolarimetry 
observations. \citet{baron2000} have presented 
model atmosphere fits to the early-time optical and HST-UV 
spectra, indicating that an enhanced He abundance is required to 
fit the data satisfactorily. They further use the very blue 
continuum of the early spectrum to determine a reddening.
The expanding photosphere method (EPM) has been applied
to SN1999em by \citet{hamuy2001} to determine 
a distance to the host galaxy of $7.5\pm0.5$\,Mpc, 
illustrating the possibility of using SNe\,II-P as luminous distance 
indicators. Chandra and radio observations of SN1999em have 
been used to probe the interaction of the SN ejecta with the 
circumstellar material, which are consistent with a 
mass-loss rate of $\sim2\times10^{-6}$\msol\,yr$^{-1}$ and a
slow moving wind velocity of 10\kms \citep{pool2001}. 
Given the substantial interest in this bright supernova and the 
extensive multi-wavelength observations of the event
it is of great interest to have direct information on the progenitor star. 
Further it would be desirable to have more detections of progenitor
stars (as in SN1987A) in order to draw a meaningful physical 
picture of what causes the different varieties of core-collapse
events. 

By chance there are optical images of this galaxy taken 7 years before
SN1999em occurred in the archive of the Canada France Hawaii
Telescope, maintained at the Canadian Astronomy Data
Centre\footnote{http//:cadcwww.dao.nrc.ca/cfht/}.  These
high-resolution images were taken by \citet{sohn98},
who presented photometry of the luminous supergiant members of the
galaxy. Amongst other results in this paper, a distance of 
7.8$\pm1$\,kpc is derived from the magnitudes of the galaxy's brightest
stars. As SNe Type\,II are thought to have luminous supergiant
progenitors, high quality pre-explosion images of nearby
galaxies which resolve the brightest stars could allow direct
detection of progenitors, or at least limits to be set on luminosity
in the event of a non-detection.  In this paper we present an accurate
astrometric determination of the position of SN1999em on the
pre-explosion frames. We show that there is no detection of a point
source at this position which could be interpreted as the progenitor.
The detection limits of the exposures are determined, allowing
bolometric luminosity limits and an upper mass limit to be determined
for the progenitor star. We present extended stellar evolutionary calculations
which follow 8-12\msol\ stars through C-burning and show that this phase
is extremely important in understanding SN progenitors in this mass range.
In the rest of this paper we refer to the
progenitor star as PSN1999em to distinguish between discussions of it
and the actual SN1999em event.


\section{Data Analysis}

\subsection{Astrometry, photometry  and sensitivity limits}
\label{data}

The galaxy NGC\,1637 was observed on 5th January 1992 on the CFHT with
the HRCam \citep{mc89}, with exposures of 900s, 750s
and 600s in $V, R_{\rm C}, I_{\rm C}$. The material is publicly
available through the CFHT archive at CADC$^{1}$.  The reduction,
analysis and multi-colour photometry of the bright stellar objects in
the field was presented by \citet[hereafter SD98]{sohn98}.
 Their limiting magnitudes for detection, defined as the
magnitude where DAOPHOT \citep{stet87} predicts errors of
$\pm0.5$ or greater, are $\sim$24.9, 24.8, 23.9 in $V, R_{\rm C},
I_{\rm C}$ respectively. These data hence probe stars brighter than
M$_{v}\simeq-4.9$, assuming the distance modulus from SD98, and their
estimates for average line of sight extinction of $A_V = 0.34$. The
image quality of the archive data are $0.7''$ FWHM in all three
bands. SD98 determined the colours for 435 objects in the frames which
are simultaneously detected in all three filters.  The CFHT HRCam used
a 1024$\times$1024 pixel Ford-Aerospace CCD mounted at prime focus,
with 18$\mu$m pixels, corresponding to $0.13''$ on the sky.

\begin{figure}
\epsscale{1.0}
\plotone{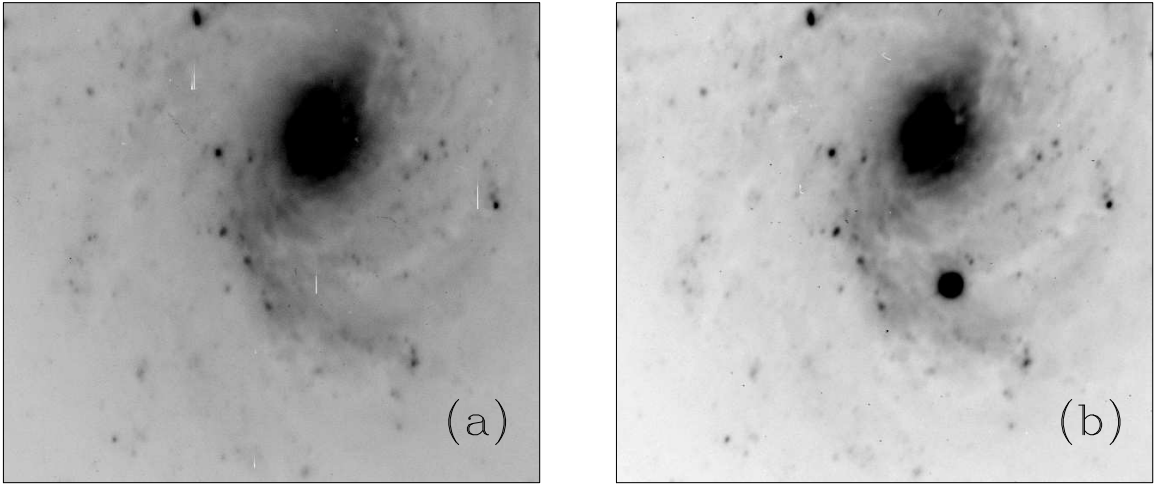}
\caption{The CFHT pre-explosion $V$-band image (a), and the 
WHT post-explosion $V$-band image (b). The FOV of both images
is $81''\times73''$, which is the full common regions available
after alignment ($0.13''$pix$^{-1}$). The similar exposure times, 
image quality and instrument characteristics result in very 
similar detection sensitivities on each frame}
\label{galaxy_images}
\end{figure}

On 28th November 1999, we obtained two $V-$band images of NGC1637 
on the William Herschel Telescope on La Palma $-$  30 days
after discovery of SN1999em. The 
AUX-port camera at Cassegrain was used, which has a 1024$^{2}$
Tektronix detector (ING CCD TEK2) at a plate scale of 
$0.11''$\,pix$^{-1}$. This was done through the ING
{\sc service} program and two exposures were taken 
(900s and 10s), during which the seeing 
was $0.7''$ FWHM. The full list of images presented 
in this paper are given in Table\,\ref{obsjournal}. 
Given the similarities between the cameras, 
telescope apertures and observing conditions in both cases, 
the sensitivities of the pre and post-explosion data 
are very similar (see Fig.\,\ref{galaxy_images}). 

\begin{figure}
\epsscale{0.45}
\plotone{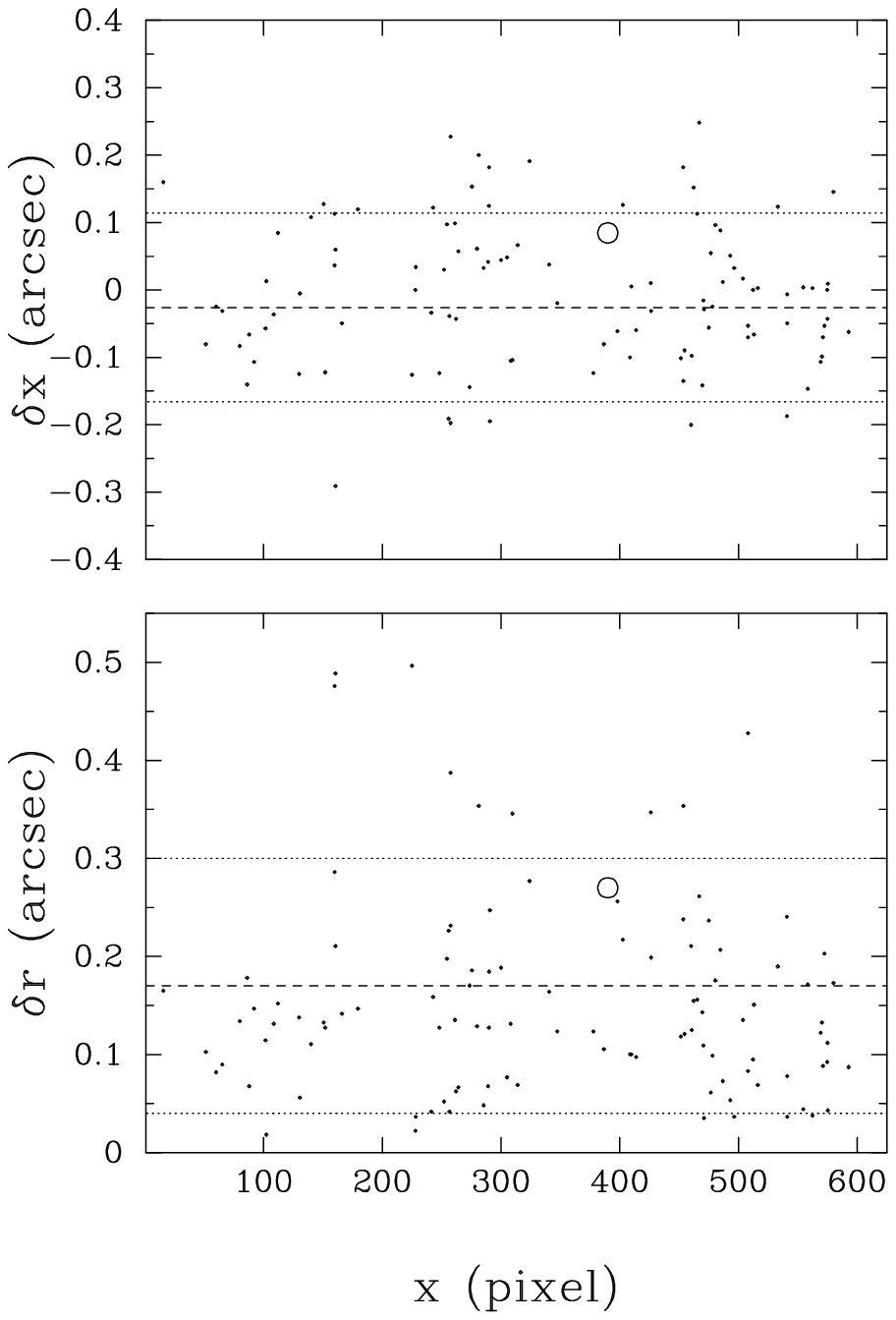}
\plotone{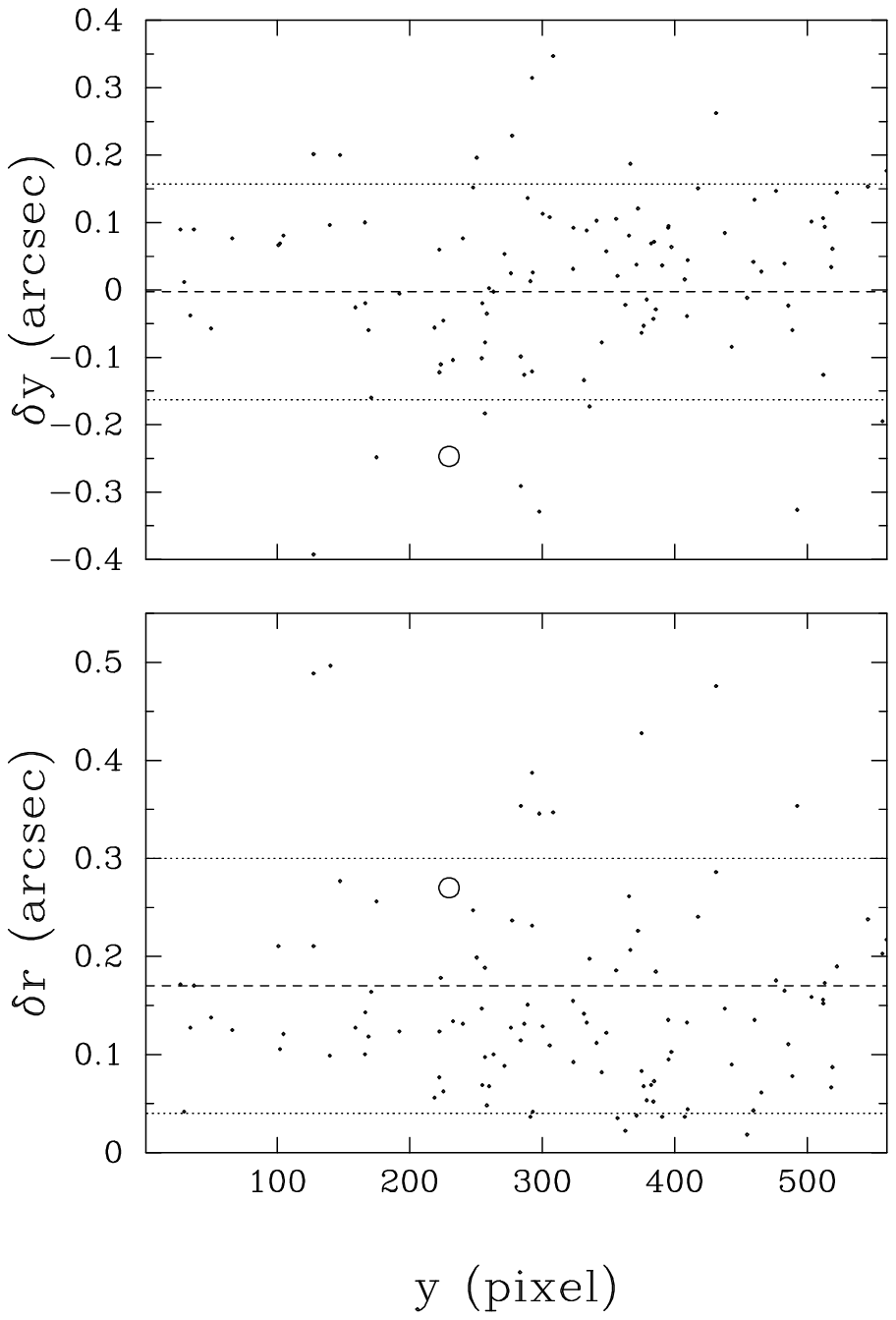}
\caption{The WHT pixel positions of the 106 paired
stars are plotted against the 
angular difference 
in their positions between the WHT and CFHT 
frames in each dimension ($\delta x$ and $\delta y$). 
Also shown is the total radial difference
between the stellar positions ($\delta$r) as a function
of $x$ and $y$ position. The difference
between SN1999em  and NGC1637-SD66 is shown as the open circle. The mean
values of the 106 star sample and the 1$\sigma$ standard deviations
are shown as the dashed dotted lines respectively.}
\label{pixel-positions}
\end{figure}

Astrometrically calibrating either of the two frames as they stand 
onto an absolute reference frame is 
not possible due to their limited FOV, and the fact that any isolated
stars outside the main body of the galaxy which could be used as
secondary astrometric standards are saturated in the deep CCD
frames. However, given the similarity in the plate scales and the
detection limits of the two data sets, we performed a simple geometric
transformation of the WHT pixel array onto the CFHT array (similar to
the method in Smartt et al. 2001a). First of all we identified ten
bright, relatively isolated stars in both the WHT 900s $V$ exposure
and CFHT $V$ frame, and measured the centroids of the stars on the WHT
frame by fitting a model point-spread-function (PSF) to each using
standard techniques in {\sc DAOPHOT} within {\sc IRAF}.  We took the
pixel coordinates of the 10 stars from the tabulated photometry of
SD98. A spatial transformation function was calculated, which fitted a
linear shift, a magnification factor and a rotation angle.  We also
tried using polynomials of various orders to fit the $x$ and $y$
mapping, but the results were no better than the simple scaling
formerly described. The transformation function was applied to the WHT
900s frame, and both were trimmed to the common region of overlap
(625$\times$560 pixels, as shown in Fig.\,\ref{galaxy_images}). As a
check on the astrometric mapping, we measured the positions of stars
which were securely identified in both the CFHT and WHT frames. We
carried out PSF-fitting photometry on the WHT frame (again using the
IRAF version of DAOPHOT), and picked relatively bright stars ($20 \la
V \la 23$) well outside the nuclear region.  We matched 106 stars on
the WHT and CFHT images, and show the differences in positions in
Fig.\,\ref{pixel-positions}.  The mean differences in the stellar
positions are
$x=-0.03'' \pm0.14$ and $y=0.00'' \pm0.16$ were the errors quoted are
standard deviations of the sample; hence there appears to be no
residual systematic difference in the pixel astrometry of these two
frames.  The mean difference in radial positions of the stars in the
CFHT and WHT frames is $\delta r = 0.17'' \pm0.13$ (see
Fig.\,\ref{pixel-positions}).

As the SN1999em was saturated in the WHT 900s frame (being
V=13.7$^{m}$), we used a second short exposure (10s) frame to
determine the centroid of the SN.  This was taken immediately after
readout of the 900s exposure, while the telescope was still guiding
smoothly.  Simply blinking the short and long exposures appeared to
show no gross shifts greater than 0.5 pixel, so we applied the
transformation function calculated for the long exposure to the 10s
frame.  Using the stellar centroid method to check for offsets between
the two frames proved problematic due to the low counts in stars in
the 10s frame; a significantly longer exposure would have led to
saturation of the SN. However we did match 5 detected stars in common
to both the 10s and the 900s exposures, and a PSF fitting of their
cores indicated mean offsets of ($-0.01''$,$0.01''$).  As there was
reasonable signal in the core region of the galaxy in the short
exposure, we cross-correlated a central portion of the transformed 10s
frame with the 900s frame. This indicated shifts of
($0.01''$,$-0.02''$), and hence we will assume that the centroid of
the SN position in the 10s frame can be directly placed on the 900s
frame with a $\pm0.03''$ error in each axes. We define the position of
SN1999em as (0,0), and all coordinates quoted hereafter are with
respect to this position (in arcseconds).

\begin{figure}
\epsscale{1.0}
\plotone{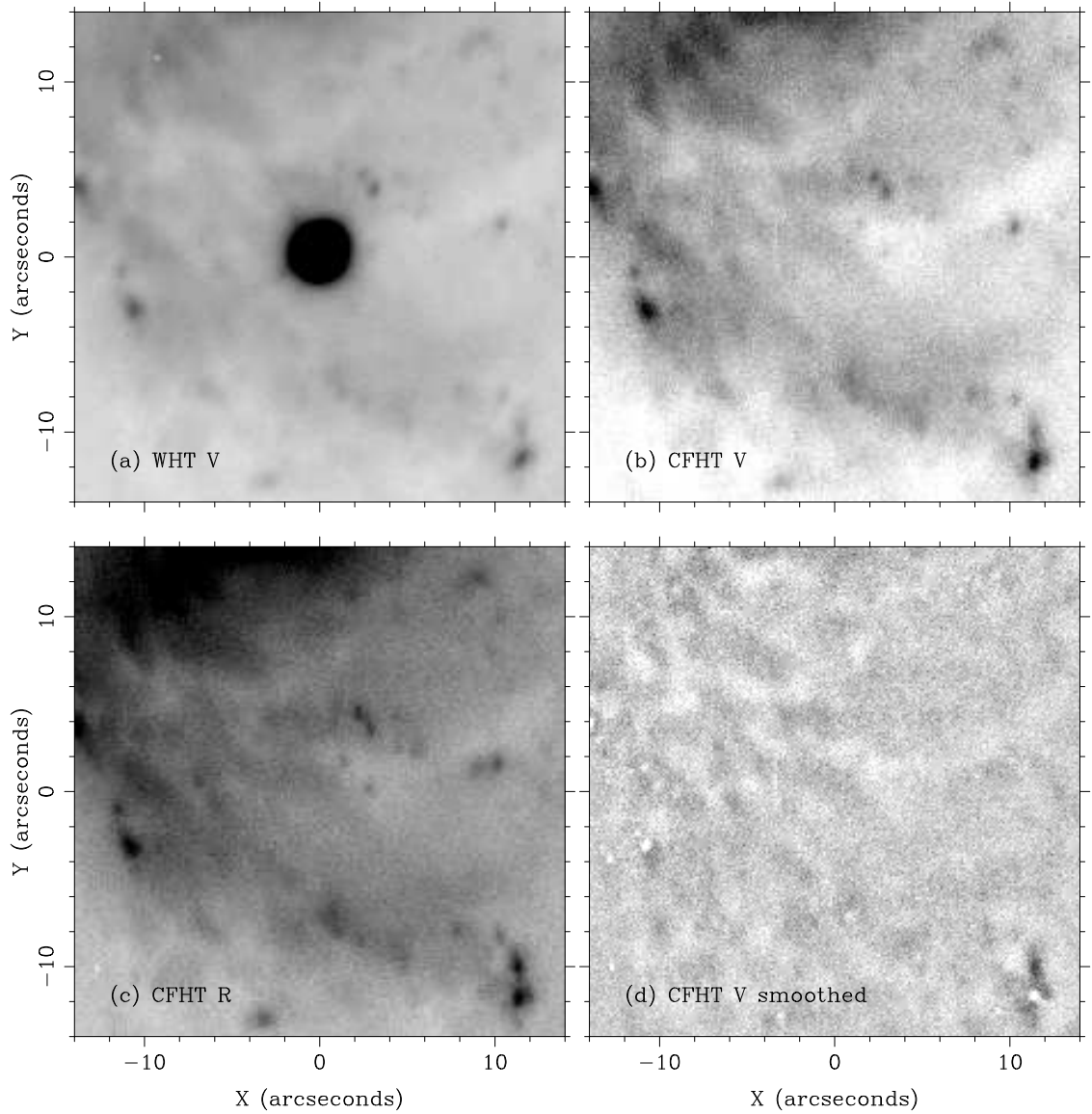}
\caption{
{\bf (a):} The position of SN1999em in the transformed WHT
post-explosion image. In this image the centroid of the SN is
saturated but a further short exposure is used to measure it
accurately, and is set at (0,0) in all frames. {\bf (b)} and {\bf
(c):} The region of the pre-explosion $VR$-band CFHT images. {\bf
(d):} An image with a smooth background removed and all PSFs from
single stars subtracted.  Sohn \& Davidge (1998) catalogue a star with
a coordinate of $(0.08'',-0.24'')$, and magnitude $V=23.47, R=23.33$
which is within the astrometric error of the transformation discussed
in Sect.\,\ref{data}. However on close inspection there is no evidence
for a point source at this position in any of the $VRI$ bands. The two
stars at (1.6, 1.9) and (2.9,0.2) are have $V=23.97, 23.15$ and
$V-R=0.26, 0.10$ respectively. The detection limit is position
dependent as the background varies considerably over small scales.}
\label{sn_closeup1}
\end{figure}

The photometry list of SD98 reports the detection of 
star \#66 (hereafter NGC1637-SD66) at 
$(0.08'',-0.24'')$ and the nearest 
other object is $2.5''$ away. Star NGC1637-SD66 is the only candidate 
for the progenitor in the existing photometry of SD98, at 
a distance of $\delta r = 0.25''$ from SN1999em. 
This does fall within the $1\sigma$ standard deviation of the 
differences in positions of the 106 matched stars, and hence
is compatible with being coincident with the supernova position 
(Fig\,\ref{pixel-positions}). 
However on closer inspection this 
does not appear to be a reliable detection of a stellar-like 
object. In Fig.\,\ref{sn_closeup1} the region around SN1999em 
is displayed from the CFHT $V$ and $R$ bands. There is no 
obvious resolved luminous object from a visual inspection
and star NGC1637-SD66 is not apparently obvious (the results for
the $I$-band data are similar).  The position 
of the supernova appears to lie on a faint ``ridge'' 
(running diagonally left-right in the figure), and the detection 
limits of the image are highly position dependent given the 
variable background. In deriving their final photometric list,  
SD98 applied a background smoothing technique to recover faint 
stars against the varying galaxy background. 
We have repeated this method to determine if any sign of
a single point source at the SN1999em position appears 
after background subtraction, following the steps  
described in \citet{sohn96}. The {\sc daophot}
package was used to fit model PSFs to
the brighter stars in the images. These were subtracted from the 
data and a boxcar median filter of pixel dimension 25$\times$25
(i.e. 5 times the seeing width)
was applied to this subtracted image. This is assumed to  
be indicative of the varying background of the galaxy and was
subtracted from the original frame. The PSF fitting routines
within {\sc daophot} were re-run on the resultant frame. The 
results from this for the $V$ band 
are shown in Fig\,\ref{sn_closeup1}(d), where
the point sources subtract off quite cleanly apart from some
objects which are not resolved but are broader than a PSF.  
Again there is no clearly identifiable point-source at the
SN1999em position after the smoothing technique is applied, 
and no object is visible in the $R$ and $I$ frames either.

We conclude that the progenitor of SN1999em is below the sensitivity
limits of the pre-explosion $VRI$ data, and that the star
NGC1637-SD66 is an erroneous detection by Sohn \& Davidge. 
Using a method similar
to \citet{smartt2001} the detection limit of these
frames can be determined in order to set constraints on the
progenitor luminosity and mass. The PSFs 
constructed for each of the $VRI$ frames were used as synthetic
stars and added into the observed frames. The detection limit is 
very dependent on position within the galaxy, because the shot noise
is background dominated as one approaches the sensitivity limit. 
Hence we only added in stars very close to the SN position. 
The SN1999em position is coincident with a bright ridge of 
material which has a significantly higher background level than 
surrounding regions. The artificial PSFs were scaled to suitable
magnitude levels (in steps of 0.2$^{m}$) and added in along this
ridge. The detection limit was defined as being when all stars
were still visible as distinct objects and had signal-levels
measured at greater than 7$\sigma$. The values are $V=23.2^m,
R=23.0^m, I=22.0^m$. These compare well with the completeness
limits determined by SD98 on the whole field of view, matching
the levels at which greater than 90\% of their artificial stars 
are recovered. 

\section{Results and discussion}

\subsection{Bolometric luminosity and mass limits for the progenitor
of SN1999em}

The detection limits on the CFHT $VRI$ data can be converted to 
limits on the bolometric magnitude and luminosity of the progenitor
object PSN1999em (as for SN1999gi in Smartt et al. 2001). 
The largest uncertainty is the reddening towards 
the progenitor. We will assume that the 
reddening derived towards SN1999em is applicable to the progenitor
line of sight. 

The reddening towards SN1999em has been estimated by two methods.
\citet{baron2000} have compared the observed early-time
spectra at two epochs with model synthetic spectra.  The non-LTE
line-blanketed model atmosphere fit to the very blue optical continuum
allows a ``temperature'' of the spectrum to be determined
simultaneously with a value for $E(B-V)$. \citet{baron2000} 
derived a best fit to the optical spectra of
T$_{\rm model} = 12000$\,K and $E(B-V)$=$0.10 \pm0.05$.  One could
conceive of increasing the temperature and also increasing the
extinction to match the continuum slope, however the Baron et
al. models illustrate that this possible degeneracy is well constrained
by the early-time blue spectrum.  A satisfactory fit to the full
spectrum (i.e.  continuum, the Ca\,{\sc ii} H$+$K, H\,{\sc i}, and
He\,{\sc i} lines) cannot be achieved with $E(B-V)$$>0.15$ and they set
this as a robust upper limit on the reddening. They also suggest that
a He abundance enhanced by a factor of 2.5 above normal is required to
fit the He\,{\sc i} $\lambda$5876 feature. A second method of reddening
estimation has been discussed in \citet{hamuy2001} which
is based on the correlation between the equivalent widths of the
Na\,{\sc i}\,D ISM lines and line of sight reddening. A calibration of
this toward Galactic OB stars has been presented by \citet{mun97}, 
and the $\sim$2\AA~ equivalent width (of the
unresolved Na\,{\sc i}\,D complex) measured by \citet{hamuy2001} 
suggests SN1999em was reddened by
$E(B-V)$$\sim1.0\pm0.15$.  This is quite discrepant from the upper
limit imposed by the model atmosphere fitting of the blue continuum. A
similar disagreement is reported by \citet{hamuy2001}
for the Type\,Ia SN1986G and both of these examples indicate the
problem with applying this Galactic calibration to derive dust
extinction towards extragalactic stars.  A closer inspection of the
\citet{mun97} results reveals that this is not
unexpected. The relatively good relation that they derive between
W(\AA) vs. $E(B-V)$ only holds for the sample of stars which are
$single$-$lined$ systems (at their spectral resolution of
R$\sim$16500) and moreover it is based on the Na\,{\sc i}\,D1 line
strength only.  At the spectral resolution used by 
\citet{hamuy2001} the Na\,{\sc i}\,D1 and D2 lines are blended
together as are any velocity shifted components of the individual
transitions.  The Fig.\,4 of \citet{mun97}
indicates that if the multiplicity of the line were unknown then the
scatter in $E(B-V)$ derived for a particular W(\AA) could be greater
than 1.0$^{m}$. Hence one requires high-resolution observations of the
Na\,{\sc i} ISM lines towards SNe before one begins trying to apply
this relation. Such data is not available at present in the
literature, and as one would expect the ISM lines to show multiple
velocity components, from the ISM in the Milky Way and NGC1673, use of
this method would not seem at all reliable in the particular case
of extragalactic objects. We will hence adopt the
\cite{baron2000} measurement of $E(B-V) = 0.1 \pm0.05$.

The upper limit to the absolute bolometric magnitude of the progenitor
can be calculated from the magnitude limit on $V$ simply through 
the relation given in Eqn.\ref{eqn1}; where the distance to NGC1637 
is assumed to be $7.5\pm0.5$\,Mpc (from Hamuy et al. 2001), and we
assume $A_V = 3.1E(B-V)$. 
\begin{equation}
\label{eqn1}
M_{\rm bol} = 5 - 5 \log d - A_V + V + BC \\
\end{equation}
As the spectral type of the progenitor is unknown, the bolometric correction
(BC) in Eqn.\ref{eqn1} determines the upper limit to the bolometric 
magnitude, 
and the values for M$_{\rm bol}$ are given as a function of spectral 
type in Table\,\ref{luminosity_limits}. The values for BC are taken 
from \citet{dril2000}. 
The other filters can be used
in a similar manner to constrain M$_{\rm bol}$, and for $R$ the 
equivalent expression is 
\begin{eqnarray}
M_{\rm bol} & =  & 5 - 5 \log d - A_R + R+ BC_R \\
            & =  & 5 - 5 \log d - 0.749A_V + R + BC + (V-R)_0 
\end{eqnarray}
Where the bolometric correction for the R-band is $BC_R = M_{\rm bol} - M_R$. 
An $M_{\rm bol}$ can similarly be derived from the I-band limit 
assuming $A_I = 0.479A_V$. The intrinsic colours $(V-R)_0$ and $(V-I)_0$
are again taken from the tabulation of \citet{dril2000}.
The $M_{\rm bol}$ values for each filter, and the corresponding 
bolometric luminosities are listed in Table\,\ref{luminosity_limits}
as a function of spectral type  
(the luminosities are calculated assuming the solar $M_{\rm bol}=4.74$). 

These luminosity limits can be compared with stellar evolutionary
tracks to determine an upper mass limit for PSN1999em.
The use of solar metallicity tracks is justified from the measurement
of nebular abundances in NGC1637.  Metallicities of H\,{\sc ii}
regions have been determined by \citet{vanzee98}
which provide measurements of the galactic abundance gradients of
oxygen, nitrogen and sulphur. The central oxygen abundance is
determined as $12 + \log (\rm O/H) = 9.18\pm0.07$ and the gradient is
$-0.07 \pm0.03$ dex\,kpc$^{-1}$. The SN occurred at a distance $23''$
from the nucleus, which would suggest an oxygen abundance of 9.1\,dex,
a factor of 2 above solar metallicity 
\citep[from the 8.83\,dex oxygen abundance in][]{grev98}. 
However between
distances of $19.5''$ and $30''$ from the centre of NGC1637, there are
six H\,{\sc ii} regions with measured abundances and the results
scatter between 8.94$-$9.30\,dex.  All of these abundance
determinations rely on the $R_{23}$ line ratio calibration and there
are no direct measurements of the electron temperatures.  By comparing
stellar and nebular abundances in the inner regions of M31, 
\citet{smartt2001b} have shown that the determination of oxygen
abundances from such a method at metallicities slightly higher than
solar may yield an overestimate by up to 0.3\,dex. 
\citet{pil2001} has recently compared the H\,{\sc ii} region
abundance determinations across the disk of M101 and found that the
$R_{23}$ method produces overestimates of oxygen abundance in the
inner regions compared to results derived from temperature sensitive
line ratios.  Given the scatter observed and the calibration
uncertainties, we will assume that a solar metallicity is applicable
to the Galactocentric distance at which SN1999em occurred.

\subsection{Theoretical Stellar Models}

The stellar models come from the most recent version of the
Eggleton evolution program (Eggleton 1971, 1972, 1973).  The equation
of state, which includes molecular hydrogen, pressure ionization and
coulomb interactions, is discussed by Pols et al. (1995).  The initial
composition is taken to be uniform with a hydrogen abundance $X =
0.7$, helium $Y = 0.28$ and metals $Z = 0.02$ with the meteoritic
mixture determined by Anders and Grevesse (1989).  Hydrogen burning is
allowed by the pp chain and the CNO cycles.  Helium burning is
explicitly included in the triple $\alpha$ reactions and reactions
with $^{12}$C, $^{14}$N and $^{16}$O along with carbon burning via
$\rm ^{12}C + ^{12}C$ only and the disintegration of $^{20}$Ne.
Other isotopes and reactions are not
explicitly followed.  Reaction rates are taken from Caughlan and Fowler 
(1988).
Opacity tables are those calculated by Iglesias,
Rogers and Wilson (1992) and Alexander and Ferguson (1994).  An
Eddington approximation (Woolley and Stibbs 1953) is used for the
surface boundary conditions at an optical depth of $\tau = 2/3$.  This
means that low-temperature atmospheres, in which convection extends
out as far as $\tau \approx 0.01$ (Baraffe et al. 1995), are not
modelled perfectly.  However the effect of this approximation on
observable quantities is not significant in this work (see for example
Kroupa and Tout 1997).

\begin{figure}
\epsscale{1.0}
\plotone{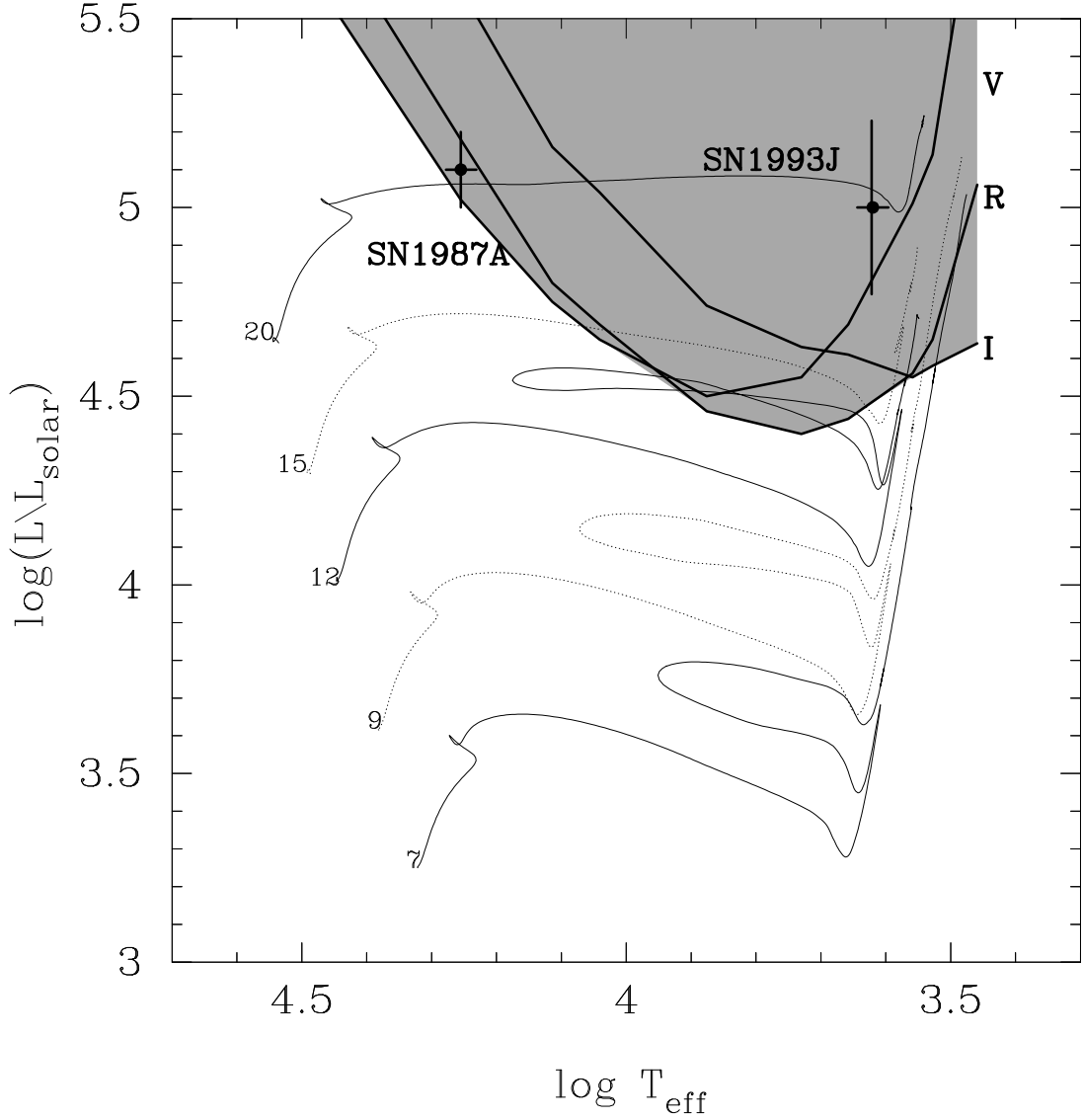}
\caption{ Evolutionary tracks
for 7$-$20\,M$_{\odot}$ stars without mass loss plotted with the
positions of the progenitor of SN1987A and SN1993J indicated.  The
tracks are alternated between dotted and solid for clarity.  The
luminosity limits on the pre-explosion progenitor of SN1999em, as a
function of stellar effective temperature, are plotted as the thick
solid lines for VRI (data from Table\,\ref{luminosity_limits}).  The
shaded area represents the region in which the progenitor would have
been detected in at least one of the filters. The progenitor of
SN1999em is unlikely to have occupied a position within this region of
the HR diagram.}
\label{psn1999em_evol}
\end{figure}

In Fig.\,\ref{psn1999em_evol} the luminosity limits are plotted with these 
evolutionary tracks.  The CFHT pre-explosion images should be sensitive to 
all objects in the shaded region of the HR diagram. Initially, 
if we were to assume that the end of core He-burning determines the 
stars pre-explosion status, we would derive an upper mass limit of 
approximately 12\msol. Stars above this mass, and corresponding luminosity,
would have $I$-band magnitudes detectable in the CFHT images. 
The error on our bolometric luminosity is $\pm$0.2\,dex, which is 
derived from combining the errors in the M$_{\rm bol}$ 
calculation in quadrature
i.e. distance ($\pm$0.5\,Mpc), reddening ($\pm0.05^m$), sensitivity
limits ($\pm0.2^{m}$), BC and intrinsic stellar colours ($\pm0.4^m$). 
We have taken a blue cutoff of B0Ia, which in effect assumes
that O-type giants or supergiants are not the precursors to SNe Type
II. There are no models to our knowledge that predict such a scenario.
However we can go further than this, as 
the final stages of evolution of stars between 7 and~$11\,M_\odot$ are 
complex and the C-burning phase plays a crucial role in 
determining their surface parameters prior to core-collapse. 
We show below that, somewhat paradoxically, the 7-11\msol\ 
stars can reach luminosities higher than their more massive counterparts
which further constrains the mass of PSN1999em. 

Nuclear evolution and mass loss compete to
determine the final course of stellar evolution, 
and dramatic differences are brought about by
changes in the order of three key events.  These are (i) the helium
exhausted core exceeding the Chandrasekhar mass, (ii) the ignition of
carbon and (iii) the second dredge-up.  A type~II supernova explodes
when the degenerate core of a star with a hydrogen envelope reaches
the Chandrasekhar mass.  It is mass loss that determines the fate of
the hydrogen but before examining the effects of its removal we shall
first consider the nuclear differences in stars of interesting masses.
\par
\begin{figure}
\epsscale{1.0}
\plotone{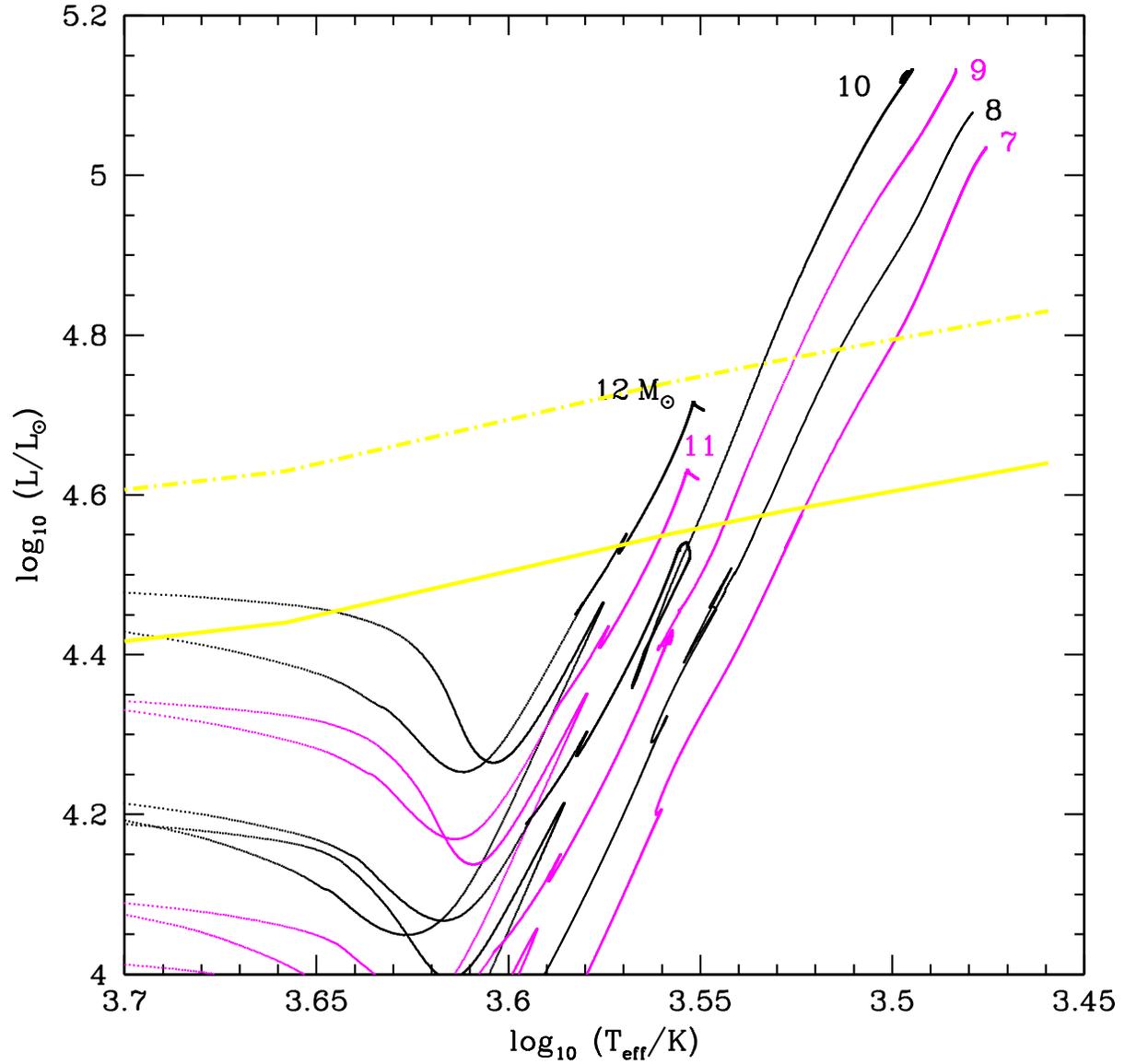}
\caption{Detail of the evolutionary tracks
for $7-12\,M_{\odot}$ stars without mass loss.  The thick solid line marks
the detection limit of the $I$-band data (as in Fig.\ref{psn1999em_evol}),  
while the thick broken line marks the maximum error in this estimate.}
\label{psn1999em_det}
\end{figure}

The higher masses are perhaps the simplest.  Above $11\,M_\odot$ the
helium burning shell passes the Chandrasekhar mass before carbon
ignites.  At this point the hydrogen burning shell lies in cooler
regions further out in the star and the luminosity is dominated by
helium burning.  Carbon burning begins non-degenerately in the core
and moves without note to shell burning leaving behind an
oxygen-neon-magnesium core (hereinafter ONe) that soon reaches $M_{\rm
Ch}$.  The entire carbon burning phase has very little effect on the star's
position in the H-R diagram. This is 
essentially the hook at the top of the 11
and~$12\,M_\odot$ tracks in Fig.~\ref{psn1999em_evol} (more clearly
seen in the close-up
of this stellar end game in Fig.~\ref{psn1999em_det}). The
supernova explodes when the star is more or less where it was when
carbon ignited.  For this reason it is efficient and often sufficient
to evolve models of such stars only to the point of carbon ignition if
we wish to determine the luminosity and effective temperature of the
immediate progenitor of the supernova.  Up to this point our tracks
are very similar to those used by others such as
the Geneva Group \citep{mey94,sch92}.  As we shall demonstrate, this
is not enough for lower masses.
\par
At the other end of this interesting region, intermediate stars such
as a $7\,M_\odot$ experience a second dredge-up before carbon
ignition.  The rate of helium burning in its shell rises so that the
hydrogen-burning shell expands, cools and goes out on the asymptotic
giant branch (hereinafter AGB).  Such a star's convective envelope is
then able to pass the extinct H-burning region and carry fresh
hydrogen down close to the He-burning shell.  A new H-burning shell
ignites and once again dominates the luminosity.  Because hydrogen is
burning at a much greater, hotter depth in the star, it can reach
considerably higher total luminosities than its more massive
counterparts that do not experience a second dredge-up.  This is again
apparent in Fig.~\ref{psn1999em_det}.  A $7\,M_\odot$ star is
expected to experience thermal pulses but
these are not of interest to us here.  Eventually it ignites carbon
but while the core mass within the He/H 
shells is still well below $M_{\rm Ch}$.  Carbon
burning moves to a shell and, without mass loss, all three burning
shells would burn out to $M_{\rm Ch}$ followed by a supernova
explosion.  However $7\,M_\odot$ stars do not empirically lead to
SNe~II: in practice stellar winds remove their H-rich envelopes
leaving ONe white dwarfs.  Our $8\,M_\odot$ star evolves similarly.
It lies close to the boundary between which mass loss and nuclear
evolution compete to end its life and may end up as an ONe white dwarf
or supernovae respectively.
\par
At intermediate masses carbon ignites before a second dredge-up but
also before the He exhausted core has reached $M_{\rm Ch}$.  This
allows for second dredge-up to take place after core carbon burning
and we do indeed find this for our 9 and~$10\,M_\odot$ stars.  Because 
of this, post-carbon-burning evolution cannot be ignored for these
stars.  As for the lower masses hydrogen once again burns in the
deeper hotter regions close to the He-burning shell.  The total
luminosity again exceeds the maximum attainable by the $12\,M_\odot$
star.  We must now wait until all three burning shells reach $M_{\rm
Ch}$ before a supernova explosion.  Because conditions deep inside are 
not much affected by envelope mass the maximum luminosity attained by
the 9 and~$10\,M_\odot$ stars are similar, the smaller envelope simply 
making the $9\,M_\odot$ redder. 
Similar behaviour was found by \citet{garcia94}. The
Geneva group's models do not show this behaviour because they have a
large convective overshoot during core helium burning that ensures
that the helium exhausted core exceeds the Chandrasekhar mass before
carbon ignition.  Our models do not include convective overshooting
though they do treat semi-convection in a consistent manner
\citep{eggleton1972}

\par
\begin{figure}
\epsscale{1.0}
\plotone{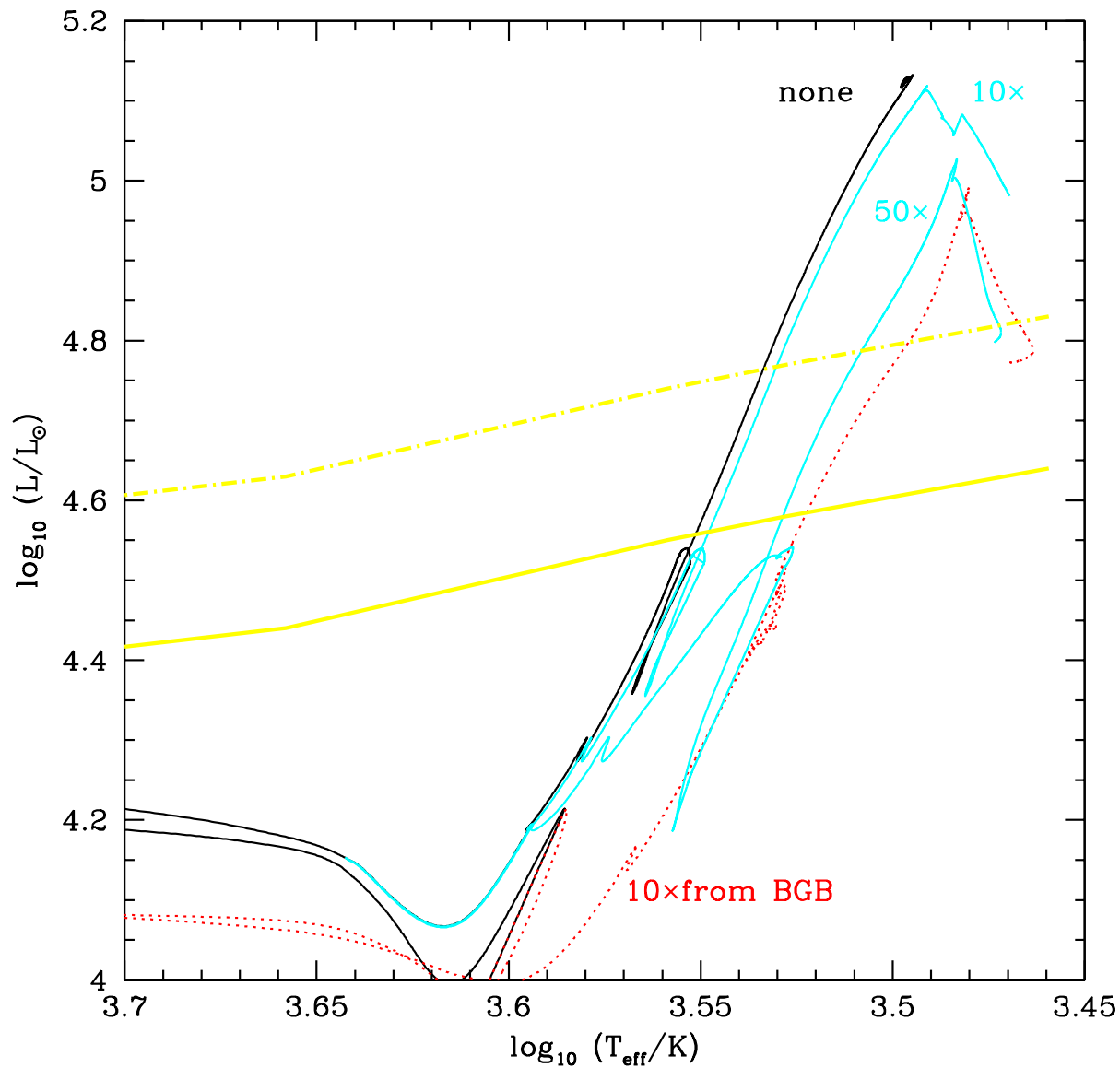}
\caption{ Detail of the final stages of evolution of 10\msol\ stars with various
mass loss rates. The 10$\times$ and 50$\times$ tracks represent 10 
and 50 times a standard Reimers mass loss rate of 
$4\times10^{-13}\frac{L}{L_\odot}\frac{R}{R_\odot}\frac{M_\odot}{M}$
\,\msol\,yr$^{-1}$, applied from the beginning of the AGB. The dotted track
has 10 times Reimers applied from the base of red giant branch. The 
observational limits imposed by the $I$-band magnitude are as in 
Fig.\ref{psn1999em_det}}
\label{psn1999em_ml}
\end{figure}

Superimposed on Fig.~\ref{psn1999em_det} is our detection limit.  Perhaps
surprisingly, all the stars lie above the limit when their cores are
due to collapse with only the $11$ and~$12\,M_\odot$ within our maximum 
error.  But we have yet
to consider mass loss.  Because the 12 and $11\,M_\odot$ stars never
get quite so luminous nor undergo thermal pulses that prolong the AGB
evolution and perhaps drive superwinds \citep{vassiliadis1993}, we do
not expect significant mass loss from these stars.  Indeed any
reasonable mass-loss rates, and we have experimented with 
\citep{reimers75} and
ten times Reimers, have negligible effect on the point of explosion in
the H-R diagram.  We have already pointed out that a $7\,M_\odot$ star
is expected to lose its envelope and end up as an ONe white dwarf.
Most of this mass loss is thought to take place on the upper AGB after
second dredge-up when mass loss dominates over nuclear evolution.  We
must therefore investigate the possible effects of mass loss on our
interestingly intermediate mass stars.  
Fig.~\ref{psn1999em_ml} illustrates a variety
of cases for a $10\,M_\odot$ star.  As with the 11 and $12\,M_\odot$
we do not expect significant mass loss before the second dredge-up
though we do allow for the possibility.  Four cases serve to
illustrate the outcomes, (i)~a Reimers' rate applied
over the AGB, (ii)~ten times Reimers over the AGB, (iii)~fifty times
Reimers over the AGB and exceptionally
(iv)~ten times Reimers from the beginning of the red giant branch.
Application of the standard Reimers' rate from the RGB has little
effect on the final outcome and typical currently favoured schemes are
well represented by less than Reimers before second dredge up rising
to ten times Reimers subsequently (e.g. Hurley et al. 2000).  When ten
times Reimers is applied from the start of the AGB nuclear evolution
continues to dominate until second dredge-up is complete and so does
not reduce the maximum luminosity achieved.  However mass loss and
nuclear burning drive the evolution at similar rates.  The star moves
to the red and its luminosity drops, as its hydrogen envelope is
stripped, until the burning shells reach $M_{\rm Ch}$.  At this
particular rate this happens when the total mass has fallen to about
$5\,M_\odot$ and the star has just passed our detection limit.
Empirically $10\,M_\odot$ stars are not expected to end up as white
dwarfs so nuclear evolution must win out somewhere along this part of
the track.  Recall further that 1999em is a type~II-P so a
substantial H-envelope must remain at the time of the explosion.
\par
At fifty times the Reimers rate the star will lose its hydrogen
envelope before explosion leaving an ONe white dwarf.  We follow it
until it turns back to the blue at a total mass of $2.5\,M_\odot$
while its core is still only $1.37\,M_\odot$.
The extreme case of rapid mass loss from the RGB onwards meets a
similar end.  Second
dredge-up still occurs after core carbon burning and high luminosities 
are still reached.  However the lower total mass, which has already
fallen to $6.4\,M_\odot$ by the start of the AGB and to $5.6\,M_\odot$ 
by the end of the second dredge-up, makes the track much redder and
beyond our detection limit.  This star inevitably ends its life as a
white dwarf and not a supernova.  At the even higher rate of one
hundred times Reimers (not shown in the figure) the star loses its
envelope before second dredge-up to become a naked helium star which
is unable to produce a type II-P supernova.
\par
With our incomplete understanding of mass loss our results remain
somewhat uncertain, but at present 
we exclude all progenitors around $8-10\,M_\odot$, even those with very
high mass loss. When we do detect a progenitor star
we shall have a lot more to say (see \ref{others_future}). 

\subsection{The nature of the progenitors PSN1999em and 
PSN1999gi}

The two supernovae SN1999em and SN1999gi are quite 
similar in their observed properties and evolution. 
They had peak absolute visual magnitudes of $-15.8$ and 
$-16.0$ respectively and both show classic plateau behaviour
(Type\,II-P). They occurred in regions of similar metallicity 
in their host galaxies; their progenitors likely had
metallicities somewhere between solar and twice solar. 
The observational limits that we have set also show that their
progenitors may have had quite similar masses. Assuming that 
stars of below 8\msol\ cannot undergo core-collapse, it 
is likely that the progenitors of these SNe II-P were in the 
range 8-12\msol. We can be reasonably confident in ruling out
very high mass stars (i.e. $>$15\msol) as progenitors, 
unless the reddening measured towards the SNe was considerably
changed by the explosions 
\citep[for example as suggested for SN1982E by][]{grah86}. 
The atmospheric parameters of the stellar models at the end of 
core C-burning are broadly consistent with observational parameters
derived from spectra and interferometric measurements of mid M-type 
supergiants \citep[M1-M2 Iab-Ia, see for example][]{white80,dekoter88}. 

The picture of PSN1999gi and PSN1999em both having progenitors in the
range M1-M2 Iab-Ia supports the current ideas on the origin of the 
Type\,II-P events. These are thought to arise in isolated stars
which still have most of their hydrogen atmosphere intact. 
\citet{ww86} discuss the ability of 
models which treat the explosion as a central point 
energy source within an extended hydrogen envelope of a red
supergiant to reproduce the important features of Type\,II-P
lightcurves. In particular the two-three month long 
plateau event is explained by the shock 
propagating through the stars' envelope,  while the 
photosphere remains at an approximately fixed
radius and temperature. Clearly the observational 
evidence we provide cannot rule out progenitors having higher 
effective temperatures and lower radii. We can simply say 
that the limits we
set are consistent with stellar evolution and Type\,II-P
explosion theories which predict core-collapse occurring in 
red (M-type) supergiants.

In \citet{smartt2001}, in addition to determining
direct limits on a late-type progenitor, 
we discussed the possibility
that PSN1999gi was originally a high mass star (in the range 
$30-60$\msol) which 
evolved into a WR star and was then below the detection limit of 
the HST WFPC2 pre-explosion images. Such a high mass, and subsequently 
shorter evolutionary lifetime would have been more compatible with the
apparent age of the OB-association within which SN1999gi occurred. 
However this is an unlikely scenario given the amount of 
hydrogen present in the spectra of the SN, and the hydrogen poor
character of WR stars in general. The range in $M_V$ spanned by 
the WR stars is large \citep[$-3$ to $-7$][]{vacca90}, 
and the least luminous stars are generally the WNE and WCE types. 
The WCE types have negligible H atmosphere abundances, and the 
WNE generally have less than 10\% H (in mass). There are no
explosion models which predict Type\,II-P behaviour for such hot
stars with relatively small radii, and hence WR stars are unlikely to 
be candidates for PSN1999gi or PSN1999em. As there are also 
no evolutionary tracks (or explosion models) which predict that
supergiants earlier than B-types are direct supernova precursors, 
there is no reason to suggest we should investigate the region hotter
than 28500K in Fig.\,\ref{psn1999em_evol}. As SN1987A 
originated from a B3\,Ia progenitor, one could feasibly suggest that 
a less luminous star ($4.5 \lesssim \log L/L_{\odot} \lesssim 5$)
of similar temperature cannot be ruled out by the limits set in 
Fig.\,\ref{psn1999em_evol}. However SN1987A was a peculiar 
event, and much of this peculiarity was probably due to its
relatively small radius, high temperature and significant 
mass loss in a previous red-supergiant phase. Also such a 
progenitor is unlikely to be able to explain the higher peak
luminosity of SN1999em.

Further indirect evidence for a red-supergiant nature of SN1999em comes
from the radio and X-ray observations of \citet{pool2001}.
The SN is detected in X-rays  ($0.4-2$\,keV and 
$2-8$\,keV), and was also detected as a faint radio source 
several times at 8.435\,Ghz. The observations have been modelled 
by assuming the X-ray and radio fluxes originate as the 
outgoing SN shock propagates into the circumstellar material 
formed by the progenitor's stellar wind. Pooley et al. 
suggest the observations are consistent with the non-radiative
interactions of the supernova ejecta with a pre-supernova
wind from a star with a mass-loss rate 
$\sim2\times10^{-6}$\msol\,yr$^{-1}$ and a wind velocity of 10\kms. 
The uncertainty quoted for $\dot{M}$ is a factor of two, 
and in good agreement with typical $\dot{M}$ measured in 
Galactic M-supergiants; for example $\alpha$\,Scorpii 
\citep[M1.5Iab][]{dekoter88} has a mass of 15\msol, 
R=650R$_\odot$, and $\dot{M}=1\times 10^{-6}$\msol\,yr$^{-1}$. 
The measured wind parameters of 
\citet{jura90} of more massive red 
supergiants (M$>20$\msol), tend to be significantly higher 
than $\sim2\times10^{-6}$\msol\,yr$^{-1}$. The 
X-ray and radio fluxes are thus not easily reconciled with 
such a high mass progenitor, supporting the direct observational limits
we have presented. Given these estimates of mass-loss, and assuming
a mass of approximately 1.6\msol~ forms a neutron star, there is 
approximalety 10\msol~  ejected in the explosion. 
This is a large enough mass of ejected material to account for a
plateau phase lasting approximately 60-100 days and hence is still 
consistent with the behaviour of SN1999em e.g. see the plateau phases
for various masses calculated 
in Arnett \citep{arnett96}. SN1999gi has also been observed in X-rays by 
Schlegel \citep{schlegel2001}, with a luminosity very similar to SN1999em. 
Again this implies $\dot{M}\sim10^{-6}$\msol\,yr$^{-1}$, supporting
the idea that PSN1999gi was also a moderate luminosity M-supergiant
with a relatively weak wind. Applying our extended theoretical models
to the bolometric luminosity limits of Smartt et al. (2001a) would 
suggest that a mass close to 8$-$9\msol would be applicable for 
SN1999gi, consistent with the limits previously derived of 
9$^{+3}_{-2}$\msol. If we again allow approximately 2\msol~ to be 
lost by mass-loss and neutron star formation, then the SN ejecta
should be around 7$^{+3}_{-2}$\msol. While the high end of this range
is large enough to sit comfortably with models of a long plateau phase, the 
lower end is discrepant \citet{arnett96}. It is important then that 
the lightcurves of these events are presented and analysed 
within the framework of current explosion models 
to check if they are consistent with the mass limits we derive. 

\subsection{Comparison with other SNe and future prospects}
\label{others_future}

As discussed above, the two Type\,II-P SNe 1999em and 1999gi were very
similar in their observed characteristics. The similar 
progenitor mass limits further suggest that the 
progenitor stars and the core-collapse explosion were quite similar. 
It supports the idea that Type\,II-P have red supergiant progenitors, 
with large radii and retain a substantial hydrogen envelope before
core-collapse. The fact that both PSN1999gi and PSN1999em are 
at the lower mass range of stars which can undergo core-collapse
may also be a suggestion that II-P SNe {\em only} occur in such 
relatively low-mass red supergiants. However the statistics of events with 
any direct clue to the nature of progenitors are quite sparse. 
Table\,\ref{sncomp} lists the only five SNe which have some direct
study of the progenitor available 
\citep[1994I does not have an upper mass estimate;][]{barth96}. 
The two brightest and best 
studied modern supernovae are SN1987A and SN1993J, and both 
were peculiar in some way. SN1987A was unusual in many ways, 
which is generally explained by the hot progenitor star which 
had a relatively small radius, was probably a red supergiant
before undergoing a blue-loop which produced multi-layer
circumstellar material \citep[see][]{mccray93}. To be consistent in 
our comparisons, we have recalculated
the upper-mass limit for 1980K, using the absolute magnitude
limit derived by \citet{thom82} and more modern stellar colours
and bolometric corrections as applied to PSN1999em and PSN1999gi. 
The metallicity of the progenitor of SN1980K 
is estimated as 0.5Z$_{\odot}$ from the 
abundance gradient presented in 
\citet[][assuming the galactocentric radius of $5.4'$]{dutil99}, 
which is interestingly similar
to that of the LMC. We have compared the 
bolometric luminosities with the 0.008Z tracks of the Geneva group, and
derive an upper mass limit of 20$^{+5}_{-5}$\msol\ assuming that the star
ends its life as a red supergiant. This assumes a distance modulus
of 28.7 \citep{buta82} and an uncertainty of $\pm$0.5 as the distance
to this galaxy is not particularly well known. 
However we note that in the 
early B-type supergiant region  the bolometric luminosity limit 
is $\lesssim5.8$. A progenitor similar to that of
SN1987A would not have been detectable on Thompson's pre-explosion plate, 
and the progenitor mass could have been considerably greater than
20\msol\ if the star was hot. 

SN1993J began showing spectra quite
typical of a Type\,II event, but then it evolved dramatically
to resemble a Ib event \citep[][and references therein]{math2000}. 
It may be that the progenitor was stripped
of much of its outer H\,{\sc i} envelope by a close companion,
consistent  with the \citet{alder94} suggestion
that an OB companion may be responsible for the excess $UB$-band
flux from the best fit K0Ia progenitor. The progenitor star was 
not of particularly high mass, so that its own radiatively driven
stellar wind is unlikely to have been capable of  
stripping of its outer envelope to approximately 0.1$-$0.6\msol.
Apart from 1999em and 1999gi, there are no other normal SNe II-P which have 
information on the progenitor stars. Although the statistics do not allow
a reliable connection between progenitor and SNe type to be made as yet,  
we propose that normal SNe II-P result from fairly low-mass ($<12$\msol) 
single stars with large radii in the M-supergiant 
phase which have undergone only moderate mass-loss. 
The X-ray and radio observations of 1999em and 1999gi support this idea. 
Higher mass progenitors
may have complicating factors such as blue-loop evolution scenarios, 
much higher mass-loss rates, multi-component circumstellar shells
due to slow dense winds from a cool supergiant phase combined with 
hot fast winds from OB-type evolutionary phases, and all of these
may be metallicity dependent. The evolution of high mass stars 
(greater than about 20\msol) is still not well understood, 
and of particular concern is the blue-red supergiant ratio in 
Local Group galaxies and how this varies with metallicity. 
There is no stellar evolution theory that can consistently 
reproduce the observed ratios and the trend of the blue-red 
ratio increasing with metallicity 
e.g. see \citet{lang95}, although the introduction
of stellar rotation may go some way to addressing this 
\citep{mm2000,mm2001}.
In a very simple minded
approach one would envisage the higher mass stars producing 
much more heterogeneous types of SNe. The evolutionary 
tracks of 9-12\msol\ stars are much less dependent on mass-loss
rates and metallicity than the higher mass ones. Our prediction that
the SNe II-P are from the low mass end of stars which undergo core-collapse
is testable with event statistics. A stellar initial mass function 
with $\phi(m) = A m^\gamma$ and $\gamma = -2.7$ implies that 50\%
of all stars with masses above 8\msol\ should have masses in the range 
$8-12$\msol, hence $\sim$50\% of all core-collapse SNe should be of 
Type\,II-P. If one increases the upper mass limit for to 15\msol\
then 66\% of all events should be II-P. A complete evaluation of the 
statistics of all known core-collapse SNe which have sub-types is
outside the scope of this paper, given the completion and bias issues 
that one would have to review and consider in detail. However as the 
numbers of SNe which have reliable sub-types continue to rise, one 
should be able to apply this test in the near future. 

We require data on more progenitors before we can be confident of the
origins of the core-collapse SNe sub-types. Prompt and frequent
multi-wavelength observations of SNe provide quite detailed
information on the explosion and circumstellar material, and by
inference on the mass-loss and envelope properties of the
progenitor. However by having high-quality archive images of SNe sites
taken {\em prior to explosion} we can set much firmer limits on the
nature of the progenitor stars. Observations of nearby spiral and
irregular galaxies within $\sim$20\,Mpc of the Milky Way allow the
massive stellar content to be resolved. Multi-band images from the
Hubble Space Telescope of all the face on spirals would be an
excellent archive for future use when SNe are discovered. In the worst
case this will allow limits to be set on the progenitor masses, as
shown here and in Smartt et al. (2001), and should lead, in some
cases, to definite identifications of progenitor stars. Already the
HST archive contains approximately 120 Sb-Sd galaxies within
$\sim$20\,Mpc which have observations of useful depth in at least 2
broad-band filters.  There are a further 130 Sb-Sd spirals with
exposures in 1 broad-band filter.  We have a Cycle\,10 HST project to
supplement the latter 130 galaxies with 2 further filters, and observe
120 more late-type spirals in three filters.  This should give a total
of $\sim$370 Sb-Sd galaxies with HST observations and this number is
steadily increasing each year, with data coming from projects with other
scientific goals.  This is supplemented with high-quality
ground-based images from the well maintained archives of the ESO, ING,
CFHT (and soon Gemini). There are various initiatives aimed at
producing combined virtual observatories which, amongst many other
applications, have the unique historical aspect which is essential to
SNe progenitor searches. One of the first of these ({\sc
astrovirtel}\footnote{http://www.stecf.org/astrovirtel}) has already
allowed us to search multi-telescope archives (HST + ESO telescopes)
and use catalogue data as search criteria (e.g. LEDA). Along with some
manual searching of the ING and CFHT archives, this suggests there are
a further 100 spirals with ground-based observations of the quality
presented here for NGC1637.  Assuming a combined SNe II/Ib/Ic rate of
$1.00\pm0.4$\,$(100{\rm yr})^{-1}(10^{10}L^{B}_{\odot})^{-1}$
\citep{capp99}, and that the galaxies in our archive
have a mean luminosity $\sim10^{10}L^{B}_{\odot}$, then one would
expect $\sim4.7\pm2$ core-collapse SNe per year in this sample. As the
field-of-view of the WFPC2 on HST will only cover an average of 50\%
of the area of the optical disk of spirals between 10-20\,Mpc, then an
estimate of the number of SNe which will have pre-explosion archive
material available is $\sim2.4\pm2$ per year. Within a period of
$3-5$\,yrs we would hence expect the statistics presented in Table\,4
to improve significantly.  This is an example of unique science to be
done with future Virtual Observatories.

\section{Conclusions}

We have presented high-resolution ground-based images of the
pre-explosion site of the Type\,II-P supernova 1999em in 
NGC1637. The position of the supernova is accurately determined using
post-explosion 
observations with a very similar optical camera. Despite the
depth and quality of the pre-explosion observations, and the fact that
they detect many bright supergiants in the host galaxy, there is no
detection of a progenitor star. The progenitor (PSN1999em) is
below the detection limit of the each of the $VRI$-band exposures. 
By determining the detection limit of each image, a limit on the 
apparent magnitude of PSN1999em in each filter can be determined, 
from which the bolometric luminosity is calculated with some
basic assumptions. By comparing this with theoretical stellar 
evolutionary tracks an upper mass limit for PSN1999em can be 
determined. We find the following results

\begin{enumerate}
\item Assuming that the reddening derived towards SN1999em is equally 
applicable to the progenitor then we initially dervive 
an upper mass limit of approximately 12\msol\  for the initial 
mass of PSN1999em. However we find that extending the evolutionary 
tracks to the end of core C-burning has a dramatic effect on the 
final luminosities of 7-10\msol\ stars, pushing them to much 
higher luminosities than the initially more massive 11-12\msol\ stars. 
These lower mass, but more luminous objects should be detectable
in our CFHT $I$-band pre-explosion frame. Hence this 
allows us to constrain the initial mass of PSN1999em to the 
narrow range 12$\pm$1\msol. 

\item Our stellar evolutionary tracks predict that a 12\msol\
star should undergo core-collapse when an M-type supergiant. The 
limits we have set are consistent with this. Further the X-ray and
radio measurements of the interaction of the SN ejecta with 
circumstellar material allow an estimate of the mass-loss
rate and wind velocity of the progenitor star. These results are 
compatible with typical observational mass-loss rates found in low-moderate
luminosity M-type supergiants.

\item SN1999em is observationally very similar to the II-P SN1999gi
in terms of its absolute peak magnitude, and spectral characteristics. 
We have set a similar mass limit on the progenitor of SN1999gi, from 
HST pre-explosion data, of 9$^{+3}_{-2}$\msol\ (Smartt et al. 2001). 
The similarity of the progenitor mass limits and the SN events 
supports the idea that they had similar progenitor masses and 
spectral types. We suggest that the homogeneous class of 
Type II-P SN could all be from intermediate mass, single, progenitors
in the M-supergiant stage which have retained their hydrogen envelope.

\item We have compared all the information currently available on the 
5 SN progenitors which either have a mass limit or spectral type
and mass estimation. The three higher mass events have heterogeneous 
types of SNe, while the two low mass events are quite similar. 
The sparse statistics prevent any definitive
conclusions, but there is a suggestion that the post-explosion evolution of
the higher mass events are influenced by pre-explosion variations in 
mass-loss rates, metallicity, 
blue loops in the HR diagram, and complex multi-component
circumstellar shells. The homogeneous normal SN II-P may come from a
lower mass population which is not so readily influenced by these parameters. 
\end{enumerate}



\acknowledgments
SJS thanks PPARC for financial support in the form of an Advanced Fellowship
award and CAT thanks Churchill 
College for a fellowship. We thank the ING 
{\sc service} programme and Ian Skillen in particular for a quick response 
to our request for an observation of SN1999em shortly after 
discovery, and use of CFHT archive data through the Guest User
facility, Canadian Astronomy Data Center, 
which is operated by the Dominion Astrophysical Observatory for
the National Research Council of Canada's Herzberg Institute of Astrophysics.
We acknowledge the {\sc astrovirtel} initiative at ESO/ST-ECF
for new software development for searching HST/ESO archives which has 
allowed the estimation of the rates of progenitor site data availability
to be done in a very quick and efficient manner.  We thank the 
referee, David Branch, for useful suggestions on the submitted version.







\clearpage
\begin{table}
\caption{Data list of images of NGC1637\label{obsjournal}}
\begin{tabular}{lllll}
\tableline 
Date           & Telescope & Filter & Exposure & Quality \\\tableline
Jan. 5 1992 & CFHT  & $V$          & 900s & $0.7''$\\
Jan. 5 1992 & CFHT  & $R_{\rm c}$  & 750s & $0.7''$\\
Jan. 5 1992 & CFHT  & $I_{\rm c}$  & 600s & $0.7''$ \\
Nov. 28 1999 & WHT & $V$      & 900s & $0.7''$ \\
Nov. 28 1999 & WHT & $V$      & 10s & $0.7''$ \\\tableline
\end{tabular}
\end{table}

\clearpage
\begin{table}
\begin{scriptsize}
\caption{Limits on the bolometric magnitudes and luminosity of the 
progenitor PSN1999em.\label{luminosity_limits}}
\begin{tabular}{llccccccccc}
\tableline
          &          &           \multicolumn{3}{c}{Limits from $M_V = -6.49$}  &  \multicolumn{3}{c}{Limits from $M_R=-6.61$}  &  \multicolumn{3}{c}{Limits from $M_I=-7.52$}\\
 Sp. Type & $T_{\rm eff}$ (K)   & BC & $M_{\rm bol}$ & $\log L/L_{\odot}$ &  $BC+ (V-R)_0$ & $M_{\rm bol}$ & $\log L/L_{\odot}$ & $BC+ (V-I)_0$ & $M_{\rm bol}$ & $\log L/L_{\odot}$\\\tableline
  B0   &  28500 &  $-$2.63 &  $-$9.11  &  5.54 &  $-$2.98 & $-$9.59 & 5.73 & $-$3.10 &  $-$10.62 & 6.14  \\
  B3   &  18000 &  $-$1.31 &  $-$7.79  &  5.02 &  $-$1.59 & $-$8.20 & 5.18 & $-$1.69 &  $-$9.21  & 5.58 \\
  B8   &  13000 &  $-$0.66 &  $-$7.14  &  4.75 &  $-$0.64 & $-$7.25 & 4.80 & $-$0.64 &  $-$8.16  & 5.16 \\
  A0   &  11000 &  $-$0.41 &  $-$6.89  &  4.65 &  $-$0.38 & $-$6.99 & 4.69 & $-$0.33 &  $-$7.85  & 5.04 \\
  F0   &  7500  &  $-$0.01 &  $-$6.50  &  4.50 &  ~~0.20  & $-$6.41 & 4.46 &  ~~0.40 &  $-$7.12  & 4.74 \\
  G0   &  5370  &  $-$0.15 &  $-$6.64  &  4.55 &  ~~0.36  & $-$6.25 & 4.40 &  ~~0.69 &  $-$6.83  & 4.63 \\
  K0   &  4550  &  $-$0.50 &  $-$6.99  &  4.69 &  ~~0.26  & $-$6.35 & 4.44 &  ~~0.74 &  $-$6.78  & 4.61 \\
  M0   &  3620  &  $-$1.29 &  $-$7.78  &  5.01 &  $-$0.06 & $-$6.67 & 4.56 &  ~~0.88 &  $-$6.64  & 4.55 \\
  M2   &  3370  &  $-$1.62 &  $-$8.11  &  5.14 &  $-$0.28 & $-$6.89 & 4.65 &  ~~0.82 &  $-$6.70  & 4.58 \\
  M5   &  2880  &  $-$3.47 &  $-$9.96  &  5.88 &  $-$1.29 & $-$7.90 & 5.06 &  ~~0.67 &  $-$6.85  & 4.64 \\
\tableline						  
\end{tabular}					  
\end{scriptsize}
\end{table}

\clearpage
\begin{table}
\caption{Comparison of all information that is currently available
from direct observations of the progenitors of core-collapse SN.
The metallicity refers to estimates for the progenitor star, and in 
the case of the four spirals this comes from the abundance gradients
and galactocentric radii of the SN. Mass refers to the {\em main-sequence} 
mass of the progenitor.\label{sncomp}}
\begin{tabular}{lllll}\tableline
SN      & Type           & Mass       & Z & Spec. Type \\ \tableline
1987A   & II peculiar    & ~~~20\msol   & 0.5Z$_{\odot}$ &  B3Ia  \\
1980K   & II-L           & $<20$\msol & 0.5Z$_{\odot}$     &  ? \\
1993J   & IIb            & ~~~17\msol & $\sim$2Z$_{\odot}$  &  K0Ia \\
1999em  & II-P           & $<12$\msol & 1$-$2Z$_{\odot}$ & M-supergiant ? \\
1999gi  & II-P           & $<9$\msol  & $\sim$2Z$_{\odot}$ & M-supergiant ? \\
\tableline
\end{tabular}
\end{table}


\clearpage

\clearpage

\clearpage

\clearpage

\clearpage

\clearpage


\begin{thebibliography}{}


\bibitem[Alexander \& Ferguson (1994)]{alexander1994}
Alexander D. R., Ferguson J. W., 1994, \apj, 437, 879
\bibitem[Anders \& Grevesse (1989)]{anders1989}
Anders E., Grevesse, N., 1989, Geochim. Cosmochim. Acta, 53, 197
\bibitem[Arnett(1996)]{arnett96}
Arnett D., 1996, ``Supernovae and Nucleosynthesis'', Princeton University
Press
\bibitem[Baraffe et al. (1995)]{baraffe1995}
Baraffe I., Chabrier G., Allard F., Hauschildt P. H., 1995, ApJ, 446, L35
\bibitem[Aldering et al.(1994)]{alder94}
Aldering G., Humphreys R.M., Richmond M., 1994, \aj, 107, 662
\bibitem[Baron et al.(2000)]{baron2000}
Baron E., et al., 2000, \aap, 545, 444
\bibitem[Barth et al.(1996)]{barth96}
Barth A., Van Dyk S.D., Filippenko A.V., Leibundgut B., Richmond M., 
1996, \aj, 111, 2047
\bibitem[Buta(1982)]{buta82}
Buta R.J., 1982, \pasp, 94, 578
\bibitem[Cappellaro et al.(1999)]{capp99}
Cappellaro E., Evans R., Turatto M., 1999, \aap, 351, 459
\bibitem[Caughlan \& Fowler (1988)]{caughlan1988}
Caughlan G. R., Fowler W. A., 1988, At. Data Nucl. Data Tables, 40, 284
\bibitem[de Koter et al.(1988)]{dekoter88}
de Koter A., de Jager C., Nieuwenhuijzen H., 1988, \aap, 100, 146
\bibitem[Drilling \& Landolt(2000)]{dril2000}
Drilling J.S., Landolt A.U., 2000, in Allen's Astrophysical Quantities, ed. A.N. Cox,  AIP Press
\bibitem[Dutil \& Roy(1999)]{dutil99}
Dutil Y., Roy J.-R., 1999, \apj, 516, 62
\bibitem[Eggleton(1971)]{eggleton1971}
Eggleton P. P., 1971, MNRAS, 151, 351
\bibitem[Eggleton(1972)]{eggleton1972}
Eggleton P. P., 1972, MNRAS, 156, 361
\bibitem[Eggleton(1973)]{eggleton1973}
Eggleton P. P., 1973, MNRAS, 163, 279
\bibitem[Filippenko(1997)]{fili97}
Filippenko A.V., 1997, \araa, 35, 309
\bibitem[Garc\'{i}a-Berro \& Iben(1994)]{garcia94}
Garc\'{i}a-Berro E., Iben I., 1994, \apj, 434, 306
\bibitem[Graham \& Meikle(1986)]{grah86}
Graham J.R., Meikle W.P.S., 1986, \mnras 221, 789
\bibitem[Grevesse \& Sauval(1998)]{grev98}
Grevesse N., Sauval A.J., 1998, Space Science Rev., 85, 161
\bibitem[Hamuy et al.(2001)]{hamuy2001}
Hamuy M., et al., 2001, \apj in press, astro-ph/0105006
\bibitem[Iglesias et al. (1992)]{iglesias1992}
Iglesias C. A., Rogers F.J., Wilson B.G., 1992, ApJ, 397, 717
\bibitem[Jura \& Kleinmann(1990)]{jura90}
Jura M., Kleinmann S.G., 1990, \apjs, 73, 769
\bibitem[Kroupa \& Tout (1997)]{kroupa1997}
Kroupa P., Tout C. A., 1997, MNRAS, 287, 402
\bibitem[Langer \& Maeder(1995)]{lang95}
Langer N., Maeder A., 1995, \aap, 295, 685
\bibitem[Leonard et al.(2001)]{leo2001}
Leonard D.C., Filippenko A.V., Ardila D.A., Brotherton M.S., 2001, \apj, 
in press, astro-ph/0009285
\bibitem[Li(1999)]{li99}
Li W.D., 1999, IAU Circ. No. 7294
\bibitem[McErlean et al.(1999)]{mcer99}
McErlean N.D., Lennon D.J., Dufton P.L., 1999, 349, 553
\bibitem[MClure et al.(1989)]{mc89}
McClure R.D., et al. 1989, \pasp 101, 1156
\bibitem[McCray(1993)]{mccray93}
McCray R., 1993, \araa, 31, 175
\bibitem[Matheson et al.(2000)]{math2000}
Matheson T., et al., 2000, \aj, 120, 1487
\bibitem[Maeder \& Meynet(2000)]{mm2000}
Maeder A., Meynet G., 2000, \araa, 38, 143
\bibitem[Maeder \& Meynet(2001)]{mm2001}
Maeder A., Meynet G., 2001, \aap, in press, astro-ph/0105051
\bibitem[Meynet et al.(1994)]{mey94}
Meynet G., Maeder A., Schaller G. Schaerer D., Charbonnel C., 1994, 
\aaps, 103, 97
\bibitem[Munari \& Zwitter(1997)]{mun97}
Munari U., Zwitter T., 1997, \aap 318, 269
\bibitem[Pilyugin(2001)]{pil2001}
Pilyugin L.S., 2001, \aap, in press, astro-ph/0105103
\bibitem[Pols et al.(1995)]{pols1995}
Pols O.R., Tout C.A., Eggleton P.P., Han Z., 1995, MNRAS 274, 964
\bibitem[Pooley et al.(2001)]{pool2001}
Pooley D., et al., 2001, astro-ph/0103196 
\bibitem[Reimers(1975)]{reimers75}
Reimers D., 1975, Mem. Soc. R. Sci. Leige 6e.Ser., 8, 369
\bibitem[Schaller et al.(1992)]{sch92}
Schaller G., Schaerer D., Meynet G., Maeder A., 1992, \aaps, 96, 269
\bibitem[Schlegel(2001)]{schlegel2001}
Schlegel E.M., 2001, \apjl, 556, L25
\bibitem[Smartt et al.(2001a)]{smartt2001}
Smartt S.J., Gilmore G.F., Trentham N., Tout C.A., Frayn C.M., 2001a, \apjl, 
556, L29
\bibitem[Smartt et al.(2001b)]{smartt2001b}
Smartt S.J., Crowther P.A., Dufton P.L., Lennon D.J., Kudritzki R.P., 
Herrero A., McCarthy J., Bresolin F., 2001, \mnras, 325, 257
\bibitem[Sohn \& Davidge(1996)]{sohn96}
Sohn Young-Jong, Davidge T.J., 1996, \aj 111, 2280
\bibitem[Sohn \& Davidge(1998)]{sohn98}
Sohn Young-Jong, Davidge T.J., 1998, \aj 116, 130 (SD98)
\bibitem[Stetson(1987)]{stet87}
Stetson P.B., 1987, \pasp 99, 191
\bibitem[Thompson(1982)]{thom82}
Thompson L.A., 1982, \apj, 257, L63
\bibitem[Vacca \& Torres-Dogden(1990)]{vacca90}
Vacca W.D., Torres-Dogden A.V., 1990, \apjs, 73, 685
\bibitem[Van Dyk et al.(1996)]{vandyk96} 
Van Dyk S.D., Hamuy M., Fillipenko A.V., 1996, \aj, 111, 2017
\bibitem[van Zee et al.(1998)]{vanzee98}
van Zee L., Salzer J.J., Haynes M.P., O'Donoghue A.A., Balonek T.J., 1998
\aj, 116, 2805
\bibitem[Vassiliadis \& Wood (1993)]{vassiliadis1993}
Vassiliadis E., Wood P. R., 1993, \apj, 413, 641
\bibitem[Walborn et al.(1989)]{wal89}
Walborn N. et al., 1989, \aap, 219, 229
\bibitem[White \& Malin(1987)]{white87}
White G.L., Malin D.F., 1987, \nat, 327, 36
\bibitem[White(1980)]{white80}
White N.M, 1980, \aj, 242,646
\bibitem[Woolley \& Stibbs (1953)]{woolley1953}
Woolley R. v. d. R., Stibbs D. W. N., 1953, The Outer Layers of a Star,
             Clarendon Press, Oxford
\bibitem[Woosley \& Weaver(1986)]{ww86}
Woosley S.E, Weaver T.A., 1986, \araa, 24, 205 
\end{thebibliography}
\end{document}